# Realising the Right to Data Portability for the Domestic Internet of Things

Dr Lachlan Urquhart[1], Neelima Sailaja, Prof Derek McAuley

Horizon, School of Computer Science, University of Nottingham

## Introduction

Bringing the new Right to Data Portability (RTDP) from an abstract legal provision in Article 20 of the EU General Data Protection Regulation (GDPR) 2016 into practice requires a greater role for the IT design community. Simply put, the RTDP seeks to empower users by giving them greater control over their personal data, enabling them to both acquire their data and then move it around, for example to a different data controller. In this paper, we focus on how IT designers can use Privacy by Design (PbD) approaches to respond to these RTDP obligations. We are particularly interested in how the RTDP plays out for the technological context of the domestic Internet of Things (IoT). By examining the legal, commercial and technical landscape around the RTDP, we can begin to unpack the practical roadblocks and opportunities ahead in implementing the right in practice.

Legally, IT designers are increasingly being called upon to engage with regulatory compliance through Article 25 of the GDPR. This provision establishes the legal obligation to do information privacy[2] by design and default for personal data driven technologies. PbD mandates creation of safeguards to satisfy the requirements of the entire GDPR and protect data subject rights.[3] This requires IT designers to build appropriate technical or organisational safeguards into the system, taking into account: *the state of the art; cost of implementation; and nature, scope and purpose of processing*.

We are particularly interested in the domestic IoT domain, where personal information is collected by physical sensors and actuators installed in socially complex, traditionally private settings [12]. Many IoT services maintain an ongoing relationship with users where their personal data is mined and analysed with the goal of providing value added, contextually appropriate, services – for example automating routine tasks like room heating management. Readings from motion, temperature or C02 sensors can be combined to make inferences, develop behavioural profiles, and make predictions about users. There are privacy implications around how such IoT derived personal data is pieced together to create models of room and building occupancy.

IoT devices often dictate how users can interact with their personal data. Seemingly mundane design decisions around supported interactions and how a system handles data (e.g. cloud or local storage) can limit control and transparency around the personal data flows. This can impact user comprehension about how their data are being used (e.g. for profiling, targeted behavioural advertising, law enforcement investigations), and accordingly impacts their agency to exercise their legal rights (e.g. how to do subject access requests or withdraw consent). Greater attention needs to be paid to how these IoT systems are designed and their associated data driven business models to foster trust in these new technologies.

---

[1] Corresponding Author: lachlan.urquhart@nottingham.ac.uk

[2] We use terminology of 'information privacy' instead of 'data protection' in this paper.
[3] Art 25 (1) GDPR "Taking into account the *state of the art, the cost of implementation and the nature, scope, context and purposes of processing* as well as the risks of varying likelihood and severity for rights and freedoms of natural persons posed by the processing, the controller shall, both at the time of the determination of the means for processing and at the time of the processing itself, implement appropriate technical and organisational measures, such as pseudonymisation, which are designed to implement data-protection principles, such as data minimisation, in an effective manner and to *integrate the necessary safeguards into the processing in order to meet the requirements of this Regulation and protect the rights of data subjects*."

Data is utilised by many stakeholders in the IoT data supply chain, often legitimised by the legal fiction of informed consent through service terms and conditions. The scope for IoT privacy harms often stem from risks around data flowing beyond appropriate contexts, without adequate user oversight. Legally, the lack of scope users have to control data flows after it is collected is a concern for ethical and sustainable growth of the emerging IoT market. A key motivation of the RTDP is redressing the dominant model of centralising data from different sources for subsequent analytics.

In this paper, we challenge the current zeitgeist that monetisation of data, with its incumbent legal obligations, is the best business model for many personal domestic IoT systems. However, im response, through a PbD approach, novel technical platforms and architectures, like personal information management systems (PIMS), we also offer new directions. PIMS support realisation of legal rights by giving users greater control over their personal data. More broadly, they provide a route to rebalancing power asymmetries between users and service providers, by disrupting emergent commercial practices of IoT services.

The paper proceeds as follows. In Part I, we introduce the RTDP, its legal nature, what it requires from IT designers whilst reflecting limitations, particularly in relation to domestic IoT. In Part II we consider regulatory challenges from the emerging domestic IoT sector and situate our arguments within the wider legal mandate of RTDP as part of doing information PbD for the IoT. In Part III we focus on technical approaches to support realisation of RTDP, namely PIMS. The state of the art in data management architectures, tools and platforms that can provide portability, increased transparency and user control over data flows are analysed. In Part IV, we bring our perspectives together to reflect on the numerous technical, legal and business barriers and opportunities that will shape implementation of the right in practice, and how the relationship may shape future IoT innovation and business models. We finish with brief conclusions about the ongoing relationship between RTDP and PbD for the IoT.

## Part I: Legal Perspective

### a) Introducing the Right to Data Portability (RTDP)

Fundamentally, Article 20 of the GDPR provides data subjects the RTDP to increase control over their personal data,[4] but before unpacking the legal nature of how it does this, further background on the GDPR and key terminology is necessary. The long-debated GDPR passed into law in 2016 and will be enforced from May 2018, when it replaces the current, "pre-Internet" Data Protection Directive 1995 (DPD). GDPR reframes and subtly updates principles, rights and responsibilities from the existing DPD but also adds several new provisions, like RTDP. To help navigate DP terminology, Table 1 (in appendix 1) provides short and full versions of key legal terms: data subject, personal data, data processing, data controller, and data processor.

Returning to the RTDP, conceptually it is two constituent rights, namely a 'right to receive' and a 'right to transmit' data[64:4–5]. With the former, data subjects have a right to obtain their data from a data controller in a structured, commonly used, interoperable,[5] and machine

---

[4] Recital 68 GDPR

[5] "Data controllers should be encouraged to develop interoperable formats that enable data portability." (Recital 68)

readable format.[6] With the latter, data subjects have a right to move data between data controllers without hindrance, or where technically feasible, have data moved directly between data controllers.[7]

The RTDP only applies to data 'concerning' the data subject and data they 'provided to' the data controller[8] and we consider each in turn.

i) ' Concerning' – Firstly, this excludes anonymised data when it does not concern the data subject, but pseudonymous data that can be linked to the subject does count[64:10]. An issue around 'data concerning the data subject' is when it relates to third parties who may not have consented to data processing (e.g. a visitor to a property captured on a smart CCTV system). The Article 29 Working Party[9] (A29 WP) Opinion on Data Portability (revised in Spring 2017)[64] argue the term should be interpreted loosely, hence telephone, messaging or VoIP call records showing third party phone numbers could still be 'concerning' the data subject and thus given to them under a portability request.

When the new data is subsequently given to a new data controller, they should not process it in ways that adversely affect the rights and freedoms of third parties e.g. for direct marketing of products and services [64:11]. Importantly, a legal basis is still necessary for processing the third-party data, as consent is likely lacking. The A29 WP argue the legitimate interests' grounds, discussed further below, can be claimed by the new data controller. This applies when they are enabling the data subject to use their personal data (including the third-party data) **solely** for 'purely personal or household activities'. This could include users gaining greater insight into their data by analysing their energy consumption or financial data, as initiatives like the UK government led 'midata' seek to do. The A29 WP suggests controllers should provide 'tools' for data subjects to select the data they are interested in and then exclude third party data[64:12]. How these measures manifest in practice for IoT will pose HCI issues in creating intuitive tools that can visualise IoT data in a comprehensible format, with adequate granularity.

ii) 'Provided by the data subject' – Again turning to A29 WP clarity, they argue this includes two categories of data. Firstly, it is '*data actively and knowingly provided by the data subject*' like email addresses, user names, passwords input into a service. Secondly, there is '*observed data that are provided by virtue of the use of the service or the device*' like location and traffic data, smart meter data, heart rate data, and search history[64:10]. Accordingly, 'provided by' needs to be interpreted widely, but it is not unlimited, and data that has been 'inferred' or 'derived' through further analysis performed by a service provider is excluded. This includes statistical or algorithmic insights for personalisation services, profiling, or user categorisations[64:11]. For IoT, this could limit what data individuals can request from service providers, for example, tracking of home occupant movements detected by a Nest learning thermostat interactions could be okay, but not the algorithmically derived heating schedule. We return to this discussion later.

---

[6] Article 20 (1) GDPR; Recital 68
[7] Article 20(2) GDPR; 'right to transmit'; this does not extend to controllers providing technically compatible processing IT infrastructure.
[8] Article 20(1) GDPR
[9] An EU data protection regulation advisory body

Lastly, the RTDP <u>only</u> relates to personal data processed by automated means (i.e. by a computer, not paper records), where the lawful basis for doing so is either:

- The data subject has given **consent**, or explicit consent (for special categories of personal data), or
- Where processing is **necessary to fulfil performance of a contract** the data subject is party to, or at their request, prior to entering a contract.[10]

In addition to only relating to data based on contracts or consent, RTDP should be applied 'without prejudice' to other rights, like the right to erasure[11] but not to "processing necessary for the performance of a task carried out in the *public interest* or in the *exercise of official authority* vested in the controller" or when it will "*adversely affect the rights and freedoms of others*".[12] These involve complex legal balancing exercises, such as determining what is in the public interest, and what constitutes adverse effects.

To delve a bit further into the contractual and consent based nature, with contractual necessity, this includes personal data processed as part of performance of a contract with the data subject like "*the titles of books purchased by an individual from an online bookstore, or the songs listened to via a music streaming services*"[64:5]. For example, in the context of domestic IoT, pressing an Amazon Dash as entering a contract to buy, and a record being kept of what is purchased and when.

With consent, showing a data subject has given consent could become harder, especially for IoT. This is because the new framing in GDPR raises the bar above the current DPD. Consent should be freely given, specific, informed and an unambiguous indication of the individual's wishes.[13] It should be provable through a record of the consent and how it was given, individuals have a right of withdrawal, and where consent is part of a bigger contract, transparency should be increased by flagging what is being consented to and by writing clearly in plain language.[14] For 'special categories of personal data' (e.g. race or ethnic origin, sex life or orientation, political beliefs, health data, genetic or biometric data) explicit consent is necessary, but what this requires above and beyond normal consent is not clear in the law.[15] In any case, an act of affirmation is critical, whereas silence, inactivity or pre-ticked boxes are not enough.[16] The inadequacies of consent are well known, due to shortcomings of terms and conditions being the dominant mechanism [81]. These GDPR conditions establish a high threshold to be met for legal consent, and IoT businesses need to work to meet these as many IoT devices process both normal and sensitive data. This ranges from health data directly obtained from wearables like heart rate trackers or political opinions from users speaking to Amazon Alexa. Indirectly, humidity sensors in smart heat alarms may indicate sex life, and smart entry systems with fingerprint recognition gather biometric data.

---

[10] Art 20 (1)(a) GDPR "the processing is based on consent pursuant to point (a) of Article 6(1) or point (a) of Article 9(2) or on a contract pursuant to point (b) of Article 6(1); and the processing is carried out by automated means." ; Art 6(1) (a) and (b) GDPR state – "*Processing shall be lawful only if and to the extent that at least one of the following applies: (a) the data subject has given consent to the processing of his or her personal data for one or more specific purposes or (b) processing is necessary for the performance of a contract to which the data subject is party or in order to take steps at the request of the data subject prior to entering into a contract;*" Art 9(2)(a) GDPR states "*the data subject has given explicit consent to the processing of those personal data for one or more specified purposes, except where Union or Member State law provide that the prohibition referred to in paragraph 1 may not be lifted by the data subject.*"
[11] Article 20(3) GDPR
[12] Art 20(3) and 20(4) GDPR
[13] Article 9(1) GDPR
[14] Article 7, GDPR
[15] Article 9 (1)
[16] Recital 32 GDPR

Ensuring appropriate consent mechanisms are in place with IoT is a design challenge, one that is elevated when consent rules are read in conjunction with GDPR stipulations on framing of information and communications between data controllers and subjects.[17] Any communication about rights, like RTDP, is to be "*concise, transparent, intelligible and easily accessible form, using clear and plain language*"[18] in writing, electronically or even orally (when requested). Wider information provision requirements in GDPR[19] mandate controllers provide extensive information to subjects, like controller identity and contact details, any intended third country data transfer, and period data will be stored (to name but a few).

The heterogeneity of IoT device interfaces (or indeed complete lack of) pose complications for delivery of adequate information and obtaining GDPR compliant explicit consent. Utilising IoT device affordances to create new interactions through delivery methods like videos, audio and feedback from gestures like hand-waving or blinking lights and sounds may redefine consent mechanisms and shift away from the dominance of form contract terms and conditions.

Whilst this discussion of consent goes beyond RTDP directly, we include it to show that despite this paper focusing on the realising data portability, in practice compliance will not be siloed and separated out to just one right at a time. Instead, the RTDP sits not in isolation, but within a wider, complex framework, the GDPR, which mandates a vast range of compliance obligations. These vary greatly from not transferring data outside of the EU to countries with inadequate protection, doing data protection impact assessments, only collecting minimal data and storing for a limited time, ensuring secure and transparent processing and many more. Hence, responding to the RTDP goes beyond technical requirements of making systems interoperable or creating APIs so data can be ported. Instead, it raises more fundamental questions about how devices and ecosystems are designed to ensure GDPR compliance, where RTDP is one cog in a big wheel. Thinking about the ongoing relationship with data controllers and how information can be communicated to end users about their rights is key. A more engaged approach with regulatory compliance during design of systems and services could foster real changes in how personal data is handled and how privacy of users is protected.

c) Wider relationship of RTDP and IoT

GDPR provisions on PbD follow the global trend of regulators towards using technology design in regulation, as seen in Canada [15], the USA [34] and Australia[78]. In Europe, the RTDP and PbD are intrinsically linked, as PbD is required to meet the GDPR requirements and protection of rights, one of which is RTDP. This requires establishment of safeguards like organisational policies, creation of certification processes and adoption of technical measures, like pseudonymisation or encryption by default.[20] The current technical state of the art, implementation costs, and the risks to rights and freedoms of individuals need to be factored in. Furthermore, by default, controllers should ensure processing is necessary and data is not shared without individuals' knowledge. Subjects should be able to determine the "*amount of personal data collected, the extent of their processing, the period of their storage*

---

[17] Art 12 as applies to Arts 15-22 – ie the right of access, to erasure, of rectification, of restriction of processing, portability, to object, or against automated decision making
[18] Art 12(1); this also applies for information to be provided under Articles 13 and 14.
[19] Article 13 and Article 14 (the information provision differs slightly depending if data is directly obtained from data subjects, or indirectly, by sharing from another controller).
[20] Recital 78 GDPR

*and their accessibility*".[21] However, the practical difficulties of doing compliance for the distributed computational world of the IoT strengthens the case for creating new IoT infrastructures and business models. In this paper we consider Personal Information Management Systems (PIMs) as a PbD tool that can technically support realisation of the RTDP. As an aside, the European Data Protection Supervisor has great optimism about the promise of PIMS as a PbD tool [33]. Instead of services aggregating and sharing with 3rd parties personal data for analysis, users store data and permit certain service providers to either access their data from, or analyse data in, their PIMS. For IoT, the case for PIMS is even greater, as they have scope to provide users with a clear point of control over who accesses their data (or who is permitted to run analytics locally on that data), for what purposes, for how long, and so forth.

A related challenge in realising the RTDP, and more broadly PbD, in practice, is that by explicitly involving designers, there is a need to support designers in their engagement with regulatory practices. Whilst IT regulation has long recognised the importance of design as a regulatory tool [51,67] situating the role of designers in practice is difficult. Currently, the lack of awareness and practical tools for doing PbD mean it is often ignored in traditional engineering practices [24]. Whilst PbD, and what it requires, may be ostensibly accessible for regulators and lawyers, for IT designers it is less clearly prescriptive [10]. Guidance on actually how to do PbD in practice is still thin on the ground [47], hence recent explorations of how to transform legal obligations into more accessible forms, like ideation cards [54] or privacy patterns [18] to prompt reflection and action.

More holistically, there are many opportunities for IT designers, as a new type of regulator, to engage with regulation and do PbD in practice [86]. Domains of IT design like privacy enhancing technologies (PETS) [14], privacy engineering [26][76] usable privacy [42] and human data interaction [57] all have methodologies and frameworks to offer. Those orientated to the HCI tradition are particularly valuable due to end user proximity, and awareness of reflecting user values in design [37]. IT designers are not currently held to the same standards of public accountability, process and transparency as state regulators, which raises questions about their legitimacy to regulate, as non-state actors. [84,87]. However, they can engage with user practices and the environment of technology deployment through design ethnographies, prototyping, co-design and participatory approaches [20]. Through end user proximity and feedback, they can become legitimate in doing PbD, responding to their regulatory role in more prospective, novel and user centric ways [84]. Importantly, whilst IT designers can regulate through mediating user interactions by design decisions and shaping user behaviour through design, precisely how they do so is important, and reflecting on the ethical as well as legal dimensions of their role is critical.

Part II: RTDP for Domestic IoT
Our drive for doing the RTDP for the IoT is motivated by three primary concerns.

- o Firstly, the RTDP seeks to empower users, which this is critical because IoT challenges the legibility of personal data flows [79]. The ambient nature of IoT data processing, the complex range of stakeholders in the IoT ecosystem, and the practical challenges in realising GDPR compliance across these different stakeholders raise challenges, for example, even if one stakeholder is compliant, others in the supply chain might not be.

---

[21] Article 25(2) GDPR

o Secondly, the RTDP offers an opportunity to make the case for new privacy preserving business models. RTDP is a stimulus for the IT design community to reflect on how to do IoT privacy in a different way. How current online platforms handle end user privacy, limiting agency of users to control their data (e.g. by aggregating data on a mass scale from different platforms into big datasets stored in cloud to be algorithmically analysed for patterns and possible value), is not the best practice blueprint for the nascent IoT industry. By considering RTDP, we can question why IoT businesses would continue running in this manner, when it will lead to considerable compliance challenges to implement the RTDP.

o Thirdly, RTDP could enable users to derive utility from their IoT devices. Instead of connecting devices to the internet, purely for the sake of it, RTDP is the first step in users taking more control over their data to derive meaning. Through fusion of data from different sources, more accurate assessment of how they use their homes or interact with different systems may be possible, not for benefit of organisations, but for theuser's own goals and needs.

Fundamentally by looking at the RTDP for the IoT, we are questioning why systems continue to be built to collect data for centralised analysis, perhaps there is a the need to find a different technical and business model. As an emerging sector, IoT businesses are well placed to take advantage of new approaches, like PIMS. They can both realise RTDP compliance obligations by providing local control and lead a new culture around data handling. In that respect, doing PbD (specifically around RTDP) for the domestic IoT through use of PIMS could be a test bed as to how new PbD strategies can be implemented to realise GDPR rights. It is critical to learn lessons from the intersection between IoT, PbD, RTDP and PIMs to formulate blueprints for protection of other GDPR rights, like the right to object, erasure, or against profiling.

That being said, we need to think a bit more about why the IoT is a complex domain for doing PbD and protecting the RTDP. To situate this trend, it is important to remember the IoT develops from a long lineage of technological visions like ubicomp [92], calm [93] and pervasive computing [73], ambient intelligence (AmI) [1] and home automation [50]. Extensive technical research has sought to engineer these respective visions, like creating seamless networking across different contexts or building device interfaces and infrastructure that is 'invisible in use' [53]. Much research has been conducted under each of these terms, hence it is it is useful to learn from stumbling blocks of these earlier technological aspirations.

Critical reflection on the utility of engineering seamless networking, invisibility in use and calm interfaces [70,80] highlights how future centric narratives often postpone addressing issues of the present, particularly user interests, which are often marginalised [9]. Technically driven smart home research, for example, has been led by high level promises of increased efficiency, convenience and comfort for users as opposed to engaging with their contextual needs, practices and routines [50,95]. For the RTDP to become a reality, much work is necessary on technically implementing portability, for example increasing interoperability through data standards. The end user is core to this shift, and any technical implementations need to keep sight of this. Working out how to technically implement portability is important, but thinking about how it can be made usable and provide utility to the end user is just as important .

With domestic IoT, the setting of the home is critical, as smart home research has shown it is a contested social space where relationships between occupants are shaped by different domestic routines, practices and hierarchies (e.g. between parents and children) [22,69,80]. The context of deployment is important and implementations of RTDP need to ensure they do not interfere with practices of the home and negatively impact occupants (e.g. creating tensions between parents who want oversight of when teens enter and leave the home through a service that tracks this, and teens who want portability in that data so their parents cannot do this) [83]. Like smart homes, the domestic IoT is not set to come into being overnight [31], hence RTDP also has to contend with interaction between legacy systems and new data driven IoT devices and services. Learning lessons from these earlier trends means putting users at the centre of IoT product and service research, development and commercialisation.

Unlike Ubicomp or AmI, IoT generally lacks a canonical technical framing, although it was originally coined in the context of tracking objects across product supply chains [7]. Defining what is or is not the IoT is complicated, and arguably unnecessary [55]. Nowadays, it is largely typified by embedding networked sensors and actuators in diverse application areas. These include smart grids, meters and domestic energy management, connected vehicles, transport infrastructure for intelligent mobility in smart cities, and wearable lifestyle devices to quantify and feed health metrics back to users. IoT is surrounded by hype and optimism, with predictions of billions of networked devices in the next ten years[17,42] where its growth can be credited to factors like cheaper devices, access to cloud computing, increasingly advanced data analytics and ubiquitous connectivity [72].

We do not advance any canonical definition here, but there is value in outlining descriptive attributes of IoT from different stakeholders and sectors. Following analysis of IoT definitions from the UK Government Office for Science [91]; EU Article 29 Working Party [6]; UN International Telecoms Union [43]; Cisco [17]; Internet Engineering Task Force [5] and Cambridge Public Policy [25], we see IoT being portrayed as:

- o Socially embedded,
- o Remotely controllable,
- o Networked devices for information sharing between people, processes and objects,
- o An ecosystem of stakeholders around the personal data e.g. third parties,
- o Physical objects with digital presence,
- o Backend computational infrastructure (e.g. cloud, databases, servers),
- o Device to device/backend communication without direct human input

From this wide framing of IoT, a range of emergent regulatory risks for end user information privacy rights can emerge (e.g. data sensed from variety of social contexts flowing globally to the cloud, with many third parties seeking access). RTDP, as a mechanism to increase control over personal data, could be a useful response to many of these concerns.

Significant user concerns and apprehension stems from adequate control of personal data. In Europe is evident from a 2015 survey of 28,000 EU citizens on attitudes to personal data protection. It shows 66% are "*concerned about not having complete control over the information they provide online*", nearly 70% think prior explicit approval is necessary before data collection and processing, and worry about data being used for purposes different from those at collection. For IoT, the repurposing of stored data, users' insufficient knowledge of data processing by physical objects, and inadequate consent or lack of control over data sharing between such objects are other big privacy concerns [6][72].

Within IoT, oversight of information flows between devices and services, the heterogeneity of device interfaces and engaging with the subtleties and nuances of deployment settings are big challenges. With the latter, IoT devices risk exposing data and disrupting domestic social hierarchies of families or fostering tensions between flatmates [52]. Approaches to empower users to control their data streams (and the implications being drawn) are important here.

As an example, consider how data from different domestic devices can be pieced together to form occupancy models:

- Increase in $CO_2$ levels (could be the cat though [79]);
- Lighting and heating control;
- Passive Infrared (PIR) sensors and door contacts (i.e. traditional sensors for burglar alarms);
- In home, or more likely access point, cameras;
- Energy monitoring indicating appliance use (e.g. a kettle is easy to detect);
- Internet traffic flows (e.g. streaming audio services).

Increasing access to such sensed data for users, through the RTDP, could enable greater literacy around how devices use their data, and lead to more creative, useful responses or applications. For example, if occupants are aware of security risks around access to energy load data, perhaps they will deliberately create white noise through remotely switching devices on and off at random times. Or with motion detection, perhaps the cat won't be allowed into the living room as it moves too much and disrupts the heating schedule, leading to steep heating bills. Or with $C0_2$ sensors linked to light dimming controls, plants can be moved between rooms more frequently to change ambience and comfort through lighting.

Clearly, through IoT, detailed inferences can be drawn about daily life and "*analysis of usage patterns in such a context is likely to reveal the inhabitants' lifestyle details, habits or choices or simply their presence at home*" [6]. Combinations of non-personal data from IoT can also create sensitive personal data (which consequently need explicit user consent under data protection law) where systems collect "*data on food purchases (fridge to supermarket system) of an individual combined with the times of day they leave the house (house sensors to alarm system) might reveal their religion*" [25]. Providing greater awareness to end users of the stories their IoT objects can tell about them could lead to questioning and renegotiation of relationships between users, their data footprint, and their devices.

## Part III: Realising the RTDP with PIMS

Having considered the nature of the right above, we now focus on some practical issues around implementation. As we've suggested, data portability is about more than the technical right of data subjects to receive their data from one data controller and to transmit it to another. It is about empowering users to exercise control and choice over how their data is handled. It seeks to create more informed users who can obtain utility from accessing their data, to disrupt the established business models of platforms locking users in and importantly, to prompt creation of alternative commercial approaches to personal data in the market[64]. Strategies to implementation then, are key.

Various approaches are advised to enable re-use and portability of data, like enabling data subjects to directly download data for insertion in personal data stores or to be held by a trusted third party. Application Programming Interfaces (APIs) have a big role to play, particularly with complex, large datasets and direct transfer between controllers [61]. Where scale or complexity of data make portability hard, controllers should provide dashboards and ensure data subjects understand the "*definition, schema and structure of the personal data*"[64:18].

In terms of creating data in structured, commonly used, interoperable and machine readable formats, very little specific guidance is provided, beyond recommendations of using common open formats like CSV [82], XML, and JSON and avoiding proprietary formats [64:18]. The focus should be on ensuring interoperability, whatever that requires, as opposed to implementing specific formats [64:17]. Nevertheless, interoperable and machine readable are terms defined in law. The former is different organisations interacting to reach mutual goals by sharing knowledge, information, business practices and exchanging data.[22] The latter is open or proprietary file formats "*structured so that software applications can easily identify, recognise and extract specific data, including individual statements of fact, and their internal structure*".[23] For structured data, extensive should be attached metadata to provide context and granularity for the data being ported, and development of interoperable standards and formats by industry [64]. Despite these high level legal requirements, technical approaches to achieving data portability are still quite open.

After a request to exercise the legal rights, communications about actions by the data controller should be sent to the data subject without undue delay, normally within 1 month.[24] Where necessary, due to complex or numerous requests, this period can be extended by two months, but data subjects need to be told of delays, with justifications, within the initial one month. If controllers do not act on requests, they need to tell the data subject why within a month of the request receipt, or risk complaints with oversight authorities, like the UK ICO, and even judicial remedies. By default, any communication about requests for action are to be provided free of charge. For manifestly unfounded, excessive or repetitive requests (to be proven by the controller) a reasonable fee for administrative costs can be requested, but equally refusal to act is an option. Controllers can also request more information from data subjects to prove their identity before sending data, an important security measure.

As mentioned, we are interested in the role new personal data processing architectures can play in supporting increased user control, to enable the right to data portability, particularly for the IoT. Accordingly, we evaluate the current technical state of the art in PIMS. To introduce them briefly, PIMS put users at the core of data processing by maintaining a proximate relationship with their data, either through local or cloud based storage. Users retain control over data sharing, oversee who accesses their data, and have increased transparency around purposes of data use. PIMS can challenge the dominant business models of users 'paying' for services with their own personal data and can enable a shift away from services aggregating and analysing data to spot patterns, trends and ultimately derive value from these[33:12]. Instead, with PIMS users can permit third parties access to their data through machine readable consent terms, exercise choice over running personal analytics services, and maintain oversight of their data through dashboards [33:13]. In Section IV we shall return to consider limitations of PIMS, but illustrate various approaches to implementation here using the examples of Mydex (and MiData), Databox, Higgins, OwnCloud, and LockerProject.

---

[22] Art 2 Decision 922/2009/EC
[23] Recital 21 Directive 2013/37/EU
[24] If the request is given electronically, the response should be electronic too, unless otherwise requested by the data subject.

*Mydex*

MyDex is a personal data store that is accessible only by the user and provides them with a range of data management functions. Mydex empowers individuals by allowing them to be active participants in a personal data enabled economy by giving greater control over management of their personal data. Users are provided with tools and services that enable them to gather, aggregate, organise, analyse and share their personal data. Mydex allows the users to authorise services access to specific sets of data and control the third parties with whom this data is shared [63]. For realising the RTDP, it has built-in measures that help moving of data into and out of the system, from granular levels to mass transfers, spreading across multiple formats and platforms [58]. They provides an API that allows for attribute exchange services that include sending data to an individual's Mydex store, allowing secure access to the store (with their consent), and developing applications that run across multiple devices, accessing personal data from users. An example MyDex application is voluntary UK midata initiative.[25] Midata is a government led and has industry backing across different sectors eg energy, banking, telecoms and it seeks to help consumers get easier access to their personal data. It aims to 'give consumers access to their transaction data in a way that is electronic, portable and safe' [36], and requires organisations, to provide mobile apps that help consumers, to use their data more effectively, for example spotting trends in their consumption habits [74].

*Databox*

The Databox is a personal networked device that allows users to regain agency of their online presence[39] through active control and management of their personal data.

Service providers wishing to process user data are required to contribute an app to the Databox "app store". Part of the app is the manifest which explicitly lists data sources required and the processing to be performed. At app install time the user needs to explicitly authorise the app to access this data, while the Databox platforms enforces the subsequent access control. All data processing is then performed by the app, which runs locally on the box , thus eliminating the need for data to be sent to organisational servers. All data processing and transactions are also recorded on a transaction log by the Databox to meet regulatory and user accountability requirements.

Thus, the Databox forms a central hub in the home for collation of diverse data sources of varying nature, hence it has significant value for increasing transparency of IoT data flows. This allows for easy portability of this data between applications and services provided on the box by competing service providers.

*Higgins*

Higgins is a cloud based approach that provides storage and a locus of control to the user to

---

manage their personal data, particularly addresses, profiles, interests, contacts, friends, affiliations etc. It allows bi-directional flow of data between the store and service providers and with the user's friends and other social connections. This is enabled through the sharing and synchronisation of attribute sets between the Higgins instance and the other party involved. The set of attributes shared is site dependant and varies according to the demands placed by service providers and user's choice. Data connections between individuals are achieved through synchronisation of attribute sets between users and their respective Higgins stores [28].

Higgins specifically mentions the capability to export data from it to RDF files, whenever required by the user. This capability, to an extent, allows for data portability by helping transfer data out of the service. But this does not satisfy the requirement of data portability completely as methods and recommendations for transfer to another platform are not mentioned. Thus, the user would always have access to personal data but might not often be able to transfer it to another service to resume accessing the value of the extracted data.

*OwnCloud*

OwnCloud is a self-hosted file sync and share server that enables data access through a web user interface, sync clients and webDAV [62]. It provides an open architecture, accessible via APIs, that supports application and plugin development to process data from the OwnCloud server, based on user choice. Owncloud also supports online social interactions by allowing affiliate users to upload files to a local server through password protected public links, helping share calendar and photos with contacts and even have video calls while deciding how to collaborate on something. While Owncloud allows for third party application integration and social interactions, it mostly supports use of simple data types like photos, status messages, hosting files etc. [46]. This restriction could pose difficulties in future personal data management and portability as applications are evolving to require diverse kinds of data which OwnCloud might not be able to support. In the IoT context in particular, it could face limitations.

*Locker Project*

Locker Project gives users the ability to control their personal data by storing them in "lockers". It is client based, with its attribute store placed on the user's personal computer as opposed to on a cloud. Services can connect and synchronise with these lockers to access and use this data [66]. Lockers can store data such as website account information, photographs, contacts etc. The Locker Project uses APIs that allow services to access the data and functionality of the system to build applications with the data. Some example API methods include, accessing a user dashboard, retrieving diary entries, and managing permissions to access data. Therefore, it supports easy data portability as users can choose what applications to share their data with and help connect to those applications by providing their data through the Locker Project.

Part IV: Discussion and Conclusions

a) Technical Barriers and Implementation Challenges for Adoption of PIMS to enable Data Portability
There are a range of barriers to adoption and implementation of these platforms. Below we consider issues around low usability, hyperbolic discounting, user trust management, data format inconsistencies, platform and policy differences, and lastly, the relational and permanent nature of data. We consider each of these in turn below.

*Low usability*

Data portability is still a distant concept to the average user often because of the highly technical nature of the subject and solutions involved. While privacy enthusiasts might be comfortable parsing CSVs and XMLs and running scripts for uploading and retrieving data through API invocation, the average user needs simple, fast and useable solutions that make the decision of porting data conceivable and practical. While there is extensive research that considers methods that make data portability a fast and practical reality, there is an explicit call to the HCI community to contribute towards measures that would make adoption of these radical new solutions seamless and organic.

*Overcoming hyperbolic discounting*

Research shows that one of the primary reasons for the privacy paradox is the low value presented by actions that ensure privacy versus the service that awaits them on the other side [11,59]. This often leads to hyperbolic discounting [3] where the users are happier skipping any extra interactions that ensure privacy. Research has shown that while interacting with technologies like personal data containers there is a very high possibility of an extension of this behaviour [45]. To ensure active user participation with such technologies, there must be more research that explicitly demonstrates the value of use of such technologies and the higher symmetry of power it offers users. This would compel them to adopt such measures into their everyday data interactions, by default.

*User trust management*

A possible challenge that faces the use of PIMs is trust users have in organisations. Users often decide what data to share with whom depending on the trust and history they have with the data controllers involved [71]. Therefore, depending upon the context, there is the possibility of users choosing to share personal data with organisations they trust rather than choosing to manage the data themselves, despite the implications this sharing might entail. Particularly, in the case of PIMS, there is the possibility of the user feeling more vulnerable to security threats by placing a big portion of their data in one single location and hence refraining from using it.

*Data format inconsistencies*

Lack of consistency in data formats used is a barrier that contributes considerably to the successful implementation of data portability measures. When transferring data from one platform to another, the retrieved data should be of a form that is acceptable by the receiving entity. Organisational differences in coding styles mean that this situation is not often achieved by default. For an average user, this would pose a hindrance in the porting process with no means to input data in the form the receiver expects. Schema agnostic data storage solutions (e.g. Higgins use of RDF) that provide APIs that enable storage and retrieval could help alleviate the problem of data interoperability to an extent but the affordances they provide to the users are yet to be studied and improved upon. Solutions like the Personal Data Lake [90] addresses the big data 3V challenge (variety, volume and velocity) by accepting and storing personal data regardless of format but while the back-end of these systems might be sharpening by the minute, the user facing side of such technologies need polishing to ensure improved adoption into users' everyday lives.

*Platform differences*

Closely linked to challenges associated with data interoperability are platform interoperability issues. Porting data between platforms triggers numerous questions directed at the user, which could lead to intimidation and poor decision making. An example of this could be UI differences between the two platforms. If the two UIs do not follow similar formats, the transfer could be arduous and even meaningless. Transferring data from an image-based platform to a text based platform could highlight several inconsistencies which could lead to a dysfunctional result. To help mitigate such situations there should be solutions that guide users to choose between compatible platforms and experiences that ease them into the shift through guided transfers that help users make decisions in a rational and confident manner.

*Policy differences*

Shadowing the nature of platform inconsistencies are policy differences between the two platforms involved. Privacy policies of organisations could vary according to their needs, goals, values, priorities and jurisdiction. Hence, the collection, use and retention of personal data would be done in different ways by different data consumers. What might be held private by one platform could be displayed publicly by another and vice versa. When making the decision to port one's data, the user should be made aware of such policy differences which could otherwise result in unexpected implications, both to the user and the organisation involved.

*Accounting for the relational nature of personal data*

Often when discussing data ownership, there is a tendency to relate a single user to a particular set of data. This is often not the case as data ownership is often multi-dimensional. Data is relational [21], it is often not associated with just one person. While a camera could be owned by a single person, a photograph taken with the camera, that involves a group of people is not necessarily just one person's data. Bringing this variable into the equation increases the complexity of the scenario by a considerable scale. Several questions are raised here. How is data portability of social data managed? Does everyone involved automatically become a stake holder to make decisions in the platform shift? How will such a complex scenario be allowed for and controlled? While currently there are no technological solutions that help answer these questions, we aim to highlight the need for solutions that respect the multi-dimensional nature of personal data ownership that calls for active participation from all subjects involved.

*Permanent nature of data*

Another challenge associated with the nature of data as a resource, is its capability to be copied, re-used and propagated indefinitely [97]. This attribute of data means that once data is received by a party, if their policy (accepted by the user) states so, they could retain the data forever, despite the user having made the decision to transfer it to a different platform. This situation portrays the need for less server side data processing and management and increased activity on the user side so that data consumers do not have to access or store personal data at any time, while always providing the services they offer. Solutions like the Databox [39] are heralding in this change but the novelty of the approach means that there are a number of technological challenges to be overcome. These include possible time delays, need for more processing power on the box etc.

### b) Legal and Commercial Dimensions of the RTDP in the Emergent IoT Industry

There are numerous legal and commercial considerations around the RTDP for the IoT. Legal scholarship frequently focuses on the relationship of RTDP with competition law [8,77,88,98]. One concern is the RTDP applies widely, to all data controllers, not just those in a dominant market position [77:349]. This puts significant compliance requirements on new SMEs/start-ups, with associated challenges for their resources of bringing new products to market whilst also being compliant and putting in place infrastructure for doing data portability transfers[77][88]. Others scholars are concerned at competition law providing redress for vendor lock-in and non-actioning of data portability requests [88]. For example [8] argues competition law is not the best route to guaranteeing data portability in Europe, as traditional metrics for measuring dominant market power in a relevant market do not always translate well to online companies. Technical feasibility is another concern, particularly the level of interoperability necessary for transfer of data without hindrance [77:340,89]. [98] looks at the importance of the RTDP for cloud providers maturing, as a mechanism to increase trust in these services long term.

We take a different direction in this article, instead considering limitations of the right and challenges IoT infrastructures can pose. We then reflect on implementation concerns, such as the nature of user control enabled by PIMs and how to ensure resilience to changes in the DP law framework, namely from Brexit. Lastly, we consider IoT market willingness (or lack of) to the RTDP and PIMS.

*RTDP Limitations*

The RTDP is somewhat narrower than it first appears. One limitation is it does not cover inferences from personal data analysis, like algorithmically or statistically derived categorisations or personalisation profiles. In the context of IoT, which functions by inferences to determine user context and provide appropriate services, this is a shortcoming. Users have power under RTDP to move the raw personal data that feeds these profiles, but cannot control or move the inferences. Concerns around profiling ordinarily stem from the lack of transparency in assumptions being drawn and the second order impacts for individuals, like being denied access to services or being subject to prejudicial treatment. [23]

Responses of the IoT industry could be either positive or negative. Positively, services could find mechanisms to provide functionality without collecting personal data, mitigating the need for users to rely on DP rights. Negatively, in that scenario, users may still be subject to profiling and the impacts that creates, but as personal data is not being processed, they cannot rely on their DP rights to exercise control through the RTDP. New technical approaches to achieve functionality can call into question application of relevant laws. To take the well-worn example of cookies, Article 5(3) of the e-Privacy Directive 2002 requires user consent for information (not just personal data) to be stored or accessed on user devices i.e. with browser cookies [48]. One response has been device fingerprinting, increasingly used to uniquely identify users to track and profile them online as an alternative to placing cookies on user devices [2,40]. However, as this can be done passively, with no information placed on the end user equipment, questions can be raised as to applicability of Article 5(3). Luckily regulatory bodies have clarified fingerprinting is covered by Art 5(3) and extends to the IoT context like smart TVs, electricity meters and in car systems [6]. However, the uncertainty where technical implementation may frustrate the intent of legal provisions has corollaries with data portability. The RTDP seeks to empower end users and provide them increased control, but in the context of IoT, omission of profiles within the right could lead to adverse effects through insufficient user control.

*Beyond RTDP*

Mentioned above, it is important to recall that the right to data portability does not exist in isolation, and there are several other new data subject rights in GDPR which will also pose significant challenges for IoT designers to implement. The Right to Erasure (better known as the Right to be Forgotten), for example, gives users a right to data deletion without delay. If user consent is withdrawn or data is no longer necessary, controllers must delete it. This right must be balanced against other rights, like freedom of expression. Similarly, the Right to Restriction allows users a right to restrict data processing. Instead of full deletion, they may choose to restrict use of their data. Restricted data can only be processed in limited circumstances. Lastly, the Right to Object means users can object to their data being processed, particularly for direct marketing, and after they object, the direct marketer must stop using their data. Systematic analysis of what these require for design go beyond the scope of this paper, but compliance with the GDPR, and doing PbD for the IoT, is far broader than just doing data portability.

*IoT and the Household Exemption*

Mentioned above, in relation to third party data, the household exemption also needs further attention. This practical legal exemption ensures individuals are not subject to the full remit of DP law where unnecessary, for example in keeping a personal address book. The scope of the exemption is contested [96], but a recent test case provided clarity for IoT devices. In Rynes (Case C-212/13), it was found that homeowners running domestic CCTV could not claim the exemption when it also captures data from public spaces, such as neighbouring gardens or a street. The homeowner is processing personal data by recording videos, and thus subject to all the obligations of a data controller under the law (the current DPR). Under GDPR, this would include providing data portability. For IoT, how then do they respond to requests from individuals walking past on the street or neighbours having a BBQ that happen to be caught on a smart home security system? When designing IoT interfaces, IT designers may need to provide tools for homeowners, as possible data controllers, to manage their obligations to any data subjects visiting nearby properties, or passing by on the street.

*Establishing IoT Data Controllers*

Linked to this is challenges in realising rights for interconnected IoT devices and services. The complexity of emerging IoT ecosystems means establishing who is the data controller, and responsible for data portability, is not always easy. In the home context, again, even if a controller can be established, transparency of personal data flows between different devices, platforms and stakeholders is low. Furthermore, there are a range of actors with vested interests in devices and services from cloud storage, mobile device manufacturers, device management apps, service providers, hardware manufacturers, third party marketing and even law enforcement. The lack of legibility of data flows to end users complicates reliance on rights, although this is not unique to data portability. Nevertheless, how are data portability requests actioned and what mechanisms are in place for practically passing these between data controllers? Do users have to request from each data controller, and if so, how can that process be made practical from a UI perspective?

*IoT Device Heterogeneity*

GDPR also establishes requirements of increased information provision to users. Again, heterogeneity of IoT devices and services mean conveying detailed, legally mandated information in a user legible manner is a design challenge to be overcome. Domestic IoT device management platforms, like Works with Nest, may have a role to play, putting in place measures to enable requests to be passed down the chain to different actors in an ecosystem. However, their terms of service often limit their liability in relation to data moving to third parties, putting onus back on users to check terms and conditions of other actors [85]. The emerging nature of the IoT market means interoperability is limited, and the motive to create standards for cooperation is still being balanced by the desire to become the dominant domestic IoT home platform. Initiatives like Hypercat recognise the need to increase interoperability and standardisation, and coupled with regulatory mandates, may foster change here [44].

*Nature of User Control in PIMS: Commodity vs Human Rights*

If control over data can be returned to users, we need to be careful about the types of approaches this enables. Terminology around ownership of personal data, or data as an asset that can be traded, needs to be addressed. 'Data ownership' suggests commoditisation and supports the narrative that data can be bought and sold. Whilst business practice may suggest this is true, in the EU, information privacy rights are inalienable human rights which cannot be lost through sale and are given to all individuals, irrespective of wealth or influence [29,65,75]. In contrast, ownership models are more popular in the US where limited sector specific data protection legislation exists with data is viewed as property [94] that subjects should be paid for use of [49]. An ownership model sees data, and the associated values, being traded away for a price. As human rights cannot be bought and sold, this position is more objectionable in Europe [41] and arguably many tensions in Europe between regulators and large US companies, stem from this ideological dichotomy in the US and Europe, as we have seen in recent cases like Google Spain (Case C-131/12) or the Schrems (C-362/14) decision. As privacy and data protection law already provide mechanisms for controlling data, like data portability, ownership models are not necessary. Nevertheless, some EU developed PIMs (e.g. Hub of all Things - HAT)[26] continue to implement functionality to enable trading of personal data with third parties.

Given the privacy risks of IoT in the home, if such sensitive data is traded, there are risks to users exposing intimate details of their life. For example, if energy consumption data was traded for income, it may negatively impact the financially and socially vulnerable, especially those dealing with fuel poverty [35]. This could create inequalities between those who can afford privacy, and those that cannot, enabled by architectures with best intentions of

---

enabling more control. Through PbD, IT designers can decide how to increase control over personal data and reflect on how PIMS increase control without moving to an ownership led model.

*Resilience to Legal Change: Brexit and the loss of the RTDP mandate*

Uncertainty around long term mandate for RTDP in the UK specifically, from EU GDPR, highlights the need to build technical solutions that are resilient to legal change. Despite the UK 2016 EU Referendum, the government has committed to enacting GDPR [32]. Nevertheless, the GDPR, as a source of digital human rights and its associated mandate for change, is at risk of being disrupted post-Brexit. GDPR is an EU legal instrument called a 'Regulation', meaning it directly applies across all EU Member states, in contrast to the current DPD which is a 'Directive', meaning it must be instantiated in domestic law, in this instance, the UK Data Protection Act 1998 (DPA). However, post-Brexit, the UK will no longer be a EU member state, meaning GDPR needs to be brought into domestic law.[27] The EU plays a critical role in setting the agenda for digital human rights[28] as it provides EU Citizens rights to private and family life, and data protection in the EU Charter of Fundamental Rights (CFR). UK Parliamentary and Human Rights bodies have feared non application of CFR post-Brexit could halt future development of EU level human rights in the UK [38] [19] and remove the formal CFR mandate for DP rights post-Brexit[68].

The White Paper on the 'Great Repeal Bill' clarifies these issues [13,27] stating the law will:

- "c*onvert EU law as it stands at the moment of exit into UK law before we leave the EU*". As the GDPR, will come into force from May 2018, it will be in place at the exit date. GDPR, like all 'EU Regulations', will be directly translated into domestic law [para 2.4] so GDPR, in its entirety should be brought in, including RTDP.
- Give no role to **Court of Justice of the EU (**CJEU) jurisprudence interpreting EU laws brought into UK law.  To ensure legal certainty "*the Bill will provide that any question as to the meaning of EU-derived law will be determined in the UK courts by reference to the CJEU's case law as it exists on the day we leave the EU.* " [para 2.14] However, the role of case law after exit day is more limited.
- Give no role to the **Charter of Fundamental Rights (CFR)**, arguing as the UK won't be an EU member state anymore, CFR wouldn't apply. They downplay the importance of the CFR, arguing its rights just mirror those found in other human rights instruments already, like the European Convention on Human Rights (ECHR) or UN Charters. However, this is not true, as CFR provides the right to data protection (Art 8) which is not in ECHR.

The jurisprudence of GDPR interpretation through case law, advisory opinions, policy changes from the European Commission are all important sources for digital rights. By leaving the EU, UK citizens' rights are at risk. For data portability, and associated architectures for implementation, the legal drivers for changing business practices to provide greater control over personal data are at risk of falling away. The EU has established high level legal standards for best practice in DP law through the GDPR. As ethical IT professionals, these should not be ignored, but lack of legal sanctions or oversight for non-compliance, makes this a harder case to sell, especially when balanced against commercial costs and asymmetries in market competition (i.e. some act ethically, others do not).

---

[27] Treaty on the Functioning of the EU Art 288 – A regulation "shall have general application. It shall be binding in its entirety and directly applicable in all member states."

[28] Eg Article 29 Working Party advice and guidance on challenges of emerging technologies; Court of Justice of the EU jurisprudence like *Schrems* challenging legalities of data flows to the US or *Google Spain* on the Right to be Forgotten.

There are a number of commercial drivers that require continued UK engagement with EU DP law and privacy standards post Brexit:

- As a precondition for market access, UK needs to consider DP law. Under GDPR Article 3(2), UK data controllers, like companies, targeting goods or services towards EU citizens need to comply with GDPR standards to obtain access to the EU market.[29]
- The UK DP regime would need to be deemed 'adequate' by the EU to ensure continued EU-UK personal data flows.[30] Given the size of the UK's global technology sector, this is an important consideration[56]. Leading DP law scholars have already argued the recently passed UK Investigatory Powers Act 2016 challenges any future proclamation of adequacy[30], and a 21 Dec 2016 Court of Justice of EU (CJEU) judgment against the UK on this matter supports this position [60]. This could see UK dealing with concerns about use of non-EU cloud services, as has been the case with use of US cloud in wake of Snowden's revelations about the NSA.
- Another alternative is negotiating a bilateral EU-US 'Privacy-Shield' type EU-UK agreement, but the US experience shows this will likely be a politically complex and lengthy process.

As is clear, despite Brexit, the UK will not be able to ignore the GDPR. For PIMS services, and RTDP technologies in the IoT sector, these concerns may shape business practices for organisations with aspirations beyond the UK borders. Like with the discussion on consent above, engaging with the complexities of how law and policy interface with technology design involves looking to broader issues, but is important to situate our discussion within the wider context.

*Market Willingness for Data Portability*

As discussed, strategically, data portability aims to increase competition in the market through emergence of new services like PIMS, different actors and ultimately to maintain growth in the digital economy.

However, to disrupt the established data driven business models, we need to consider wider motivations for businesses to change to provide users more control. User concerns over data controls already exist, as discussed above, yet the privacy paradox persists and users continue to pay for services with personal data, being separated from their data at point of collection, trading privacy for convenience. Post-Snowden, trust concerns have grown in personal data infrastructures underpinning popular services [4], hence business models increasing user control, instead of the default cloud based model, could have a socially sustainable unique selling point.

---

[29] Art 3(2) GDPR
[30] Article 25 of DPD 1995; Article 45 GDPR

However, more likely to foster change however is fear of legal sanctions, where for GDPR non-compliance fines are considerable, the higher of €20m or 4% of global turnover.[31] However, in terms of actual enforcement, resource strapped regulators are balancing between stifling growth in an emerging market, like IoT, and protecting users. The risks of user harms are high in the emerging IoT market, compounded as many IoT firms are not IT firms, but manufacturers of other 'things' like locks or toys. They may lack familiarity with relevant regulatory frameworks, such as mandated good privacy and security practices, which can create vulnerabilities from ordinarily addressable security vulnerabilities e.g. hashed passwords, default encryption of communications and stored data [16].

Giving users control over their data is a big step in the right direction. However, in the long-term sustaining shifts away from current business models requires viable use cases where utility and value for end users can be derived from this additional control. As with other technology trends, like open data or smart city dashboards, applications that convey the true value take time to develop. Establishing and responding to actual user need, as we learned from predecessor technologies to IoT, should be a key driver. Data portability opens many opportunities to disrupt established digital economy business models, and we now conclude with a few brief points.

c) Opportunities for the Future and Conclusions

There is much work to be done between the IT design and legal community to realise RTDP and for the IoT in practice. As we can see above, many unanswered questions (and thus opportunities) exist around RTDP. Developing more usable interfaces for PIMS and creating dashboards that visualise datasets to users in a legible manner is one dimensions. Creating new interactions and approaches for delivering complex information in communications between users and controllers, despite device heterogeneity, is another.

However, we would like to conclude by asking, given the challenges the dominant personal data driven IoT business models create, why continue with these? Why not create new business models and architectures where data portability is less likely to be needed in the first place. Instead of hoovering up large volumes of personal data to later (hope to) derive value from the data, instead sample small volumes of the data (with sufficient detail and granularity), from legally consenting users, to understand where the value might lie.

Instead of moving personal data to the cloud, separating from the user, perform local analysis where it never leaves the device, and provide third parties access to statistical results. Changes in mindset could help data controllers avoid challenges of implementing data portability and many of the wider compliance challenges that GDPR, whilst finding innovative approaches. Data portability has scope to address many problems of users needing greater control, but more fundamentally, business models need to change so we never need to use this right in the first place.

Acknowledgments
This research is funded by the RCUK funded Horizon Digital Economy Research grant (EP/G065802/1).

---

[31] Article 83, GDPR, 2016

| Legal Term | Short Version | Full Version (incl. Relevant GDPR Article) | Examples (non exhaustive) |
|---|---|---|---|
| Data Subject | An identified or identifiable natural person. | Art 4(1) GDPR: "… an identifiable natural person is one who can be identified, directly or indirectly, in particular by reference to an identifier such as a name, an identification number, location data, an online identifier or to one or more factors specific to the physical, physiological, genetic, mental, economic, cultural or social identity of that natural person." | Individual end user who is a service user; third party bystander eg visitor to home |
| Personal Data | Information that directly or indirectly relates to the data subject | Art 4 (1) GDPR: "Any information relating to a data subject" | Name, email address, bank account number used to sign up for service account, IP or MAC address, location data tied to individual from GPS or device fingerprinting, energy consumption data from smart meter or thermostat linked to individual account, RFID data from registered smart transport cards, biometric data like fingerprints, heart rate, facial recognition profile. |
| Data Processing | Any use of, or operation performed on personal data such as collection, | Art 4(2) GDPR: any operation or set of operations which is performed on personal data or on sets of personal data, whether or not by automated means, such as collection, recording, organisation, structuring, storage, | A smart security camera recording images of an identifiable individual's face; a wearable |
| | recording, structuring, alteration and storage of data | adaptation or alteration, retrieval, consultation, use, disclosure by transmission, dissemination or otherwise making available, alignment or combination, restriction, erasure or destruction ." | device storing user details either locally or in cloud |
| Data Controller | Actors who determine the purposes of collection, and how personal data is processed | Art 4(7) GDPR: "the natural or legal person, public authority, agency or other body which, alone or jointly with others, determines the purposes and means of the processing of personal data." | Businesses across IoT supply chain eg IoT device manufacturers, IoT device management platforms, mobile app developers, social media services. |
| Data Processor | Actors who process personal data on behalf of the controller | Art 4(8): "a natural or legal person, public authority, agency or other body which processes personal data on behalf of the controller." | Cloud providers; outsourced database management; |

## References

1. E Aarts and S Marzano. 2003. *The New Everyday: Views on Ambient Intelligence*. 010 Publishers, Rotterdam, Netherlands.
2. Gunes Acar, Christian Eubank, Steven Englehardt, Marc Juarez, Arvind Narayanan, and Claudia Diaz. 2014. The Web Never Forgets: Persistent Tracking Mechanisms in the Wild. *Proceedings of the 2014 ACM SIGSAC Conference on Computer and Communications Security - CCS '14*, ACM Press, 674–689.
3. Alessandro Acquisti and Jens Grossklags. 2004. Privacy attitudes and privacy behavior. *Economics of Information Security*: 165–178.
4. Andrew A. Adams, Kiyoshi Murata, Yasunori Fukuta, Yohko Orito, and Ana María Lara Palma. 2016. The view from the gallery. *ACM SIGCAS Computers and Society* 45, 3: 376–383.
5. J Arkko. 2015. *IETF RC 7452: Architectural Considerations in Smart Object Networking*. Fremont.
6. Article 29 Data Protection Working Party WP 223. 2014. Opinion 8 / 2014 on Recent Developments on the Internet of Things. *Brussels: European Commission* 23, September: 1–24.
7. Kevin Ashton. 2009. That 'Internet of Things' Thing. *RFID Journal*. Retrieved from http://www.rfidjournal.com/articles/view?4986.
8. Barbara Van der Auwermeulen. 2017. How to attribute the right to data portability in Europe: A comparative analysis of legislations. *Computer Law & Security Review* 33, 1: 57–72.

9.  G Bell and P Dourish. 2006. Yesterday's Tomorrow's: Notes on Ubiquitous Computing's Dominant Vision. *Personal and Ubiquitous Computing* 11, 2: 133–143.

10. Michael Birnhack, Eran Toch, and Irit Hadar. 2014. Privacy mindset, technological mindset. *Jurimetrics: Journal of Loaw, Science & Technology* 55: 1–71.

11. Laura Brandimarte, Alessandro Acquisti, and George Loewenstein. 2012. Misplaced Confidences: Privacy and the Control Paradox. *Social Psychological and Personality Science* 4, 3: 340–347.

12. Ian Brown. 2015. *GSR Discussion Paper Regulation and the Internet of Things*. Geneva.

13. Jack Caird Simson. 2017. *Legislating for Brexit: the Great Repeal Bill*. London.

14. L. Jean Camp and Carlos A. Osorio. 2003. Privacy-Enhancing Technologies For Internet Commerce. In O. Petrovic, M. Ksela, M. Fallenbock, and C. Kittl, eds., *Trust in the Network Economy*. Springer Verlag, Berlin, 317–331.

15. Ann Cavoukian. 2009. Privacy by Design - The 7 foundational principles - Implementation and mapping of fair information practices. *Information and Privacy Commissioner of Ontario, Canada*.

16. Richard Chirgwin. 2017. Two million recordings of families imperiled by cloud-connected toys' crappy MongoDB. *The Register*. Retrieved 3 March 2017 from https://www.theregister.co.uk/2017/02/28/cloudpets_database_leak/.

17. Cisco. 2013. *The Internet of Everything*. San Jose.

18. M Colesky, J Hoepman, and C Hillen. 2016. Critical Analysis of Privacy Design Strategies. *International Workshop on Privacy Engineering – IWPE'16*: 33–40.

19. Scottish Human Rights Commission. 2016. *Submission: Report to UN Committee on the Rights of the Child*. Edinburgh.

20. A Crabtree, P Tolmie, and Mark Rouncefield. 2012. *Doing Design Ethnography*. Springer Verlag, London.

21. Andy Crabtree and Richard Mortier. 2015. Human Data Interaction : Historical Lessons from Social Studies and CSCW. *Proceedings of the 2015 14th European Conference on Computer-Supported Cooperative Work, ECSCW'15*, 3–21.

22. Andy Crabtree and Tom Rodden. 2004. Domestic routines and design for the home. *Computer Supported Cooperative Work: CSCW: An International Journal* 13, 2: 191–220.

23. Bart Custers, Toon Calders, Bart Schermer, and Tal Zarsky. 2014. *Discrimination and Privacy in the Information Society Data Mining and Profiling in Large Databases*. Springer Berlin.

24. George Danezis, Josep Domingo-Ferrer, Marit Hansen, et al. 2014. *Privacy and data protection by design - from policy to engineering*. Heraklion.

25. Simon Deakin. 2015. *The Internet of Things: Shaping Our Future,*. Cambridge.

26. M Dennedy, J Fox, and T Finneran. 2014. *Privacy Engineer's Manifesto*. Apress, New York.

27. Department for Exiting the EU. 2017. *Legislating for the United Kingdom's withdrawal from the European Union*. London.

28. Eclipse Wiki. 2009. Higgins Personal Data Service. .

29. Lilian Edwards. 2009. Consumer Privacy Law 1: Online Direct Marketing. In *Edwards, & C. Waelde (Eds.), Law and the Internet. 3rd ed.* 489–510.

30. Lilian Edwards. 2016. Brexit: 'You don't know what you've got till it's gone'. *SCRIPTed* 13, 2: 112–117.

31. W.K. Edwards and R.E. Grinter. 2001. At Home with Ubiquitous Computing: Seven Challenges. *International Conference on Ubiquitous Computing*, 256–272.

32. Elizabeth Denham. 2016. How the ICO will be supporting the implementation of the

GDPR. *ICO*. Retrieved 3 March 2017 from
https://iconewsblog.wordpress.com/2016/10/31/how-the-ico-will-be-supporting-the-implementation-of-the-gdpr/.

33. European Data Protection Supervisor. 2016. *Opinion on Personal Information Management Systems Towards more user empowerment in managing and processing personal data*. Brussels.

34. Federal Trade Commision (FTC). 2012. Protecting Consumer in an Era of Rapid Change: Recommendations for businesses and policymakers. *Federal Trade Commision* March: 1–112.

35. Joel E. Fischer, Enrico Costanza, Sarvapali D. Ramchurn, James Colley, and Tom Rodden. 2014. Energy advisors at work. *Proceedings of the 2014 ACM International Joint Conference on Pervasive and Ubiquitous Computing - UbiComp '14 Adjunct*, ACM Press, 447–458.

36. Department for Business, Innovation and Skills. 2013. *midata Privacy Impact Assessment Report*. London.

37. Batya Friedman, Peter H Kahn, and Alan Borning. 2008. Value Sensitive Design and Information Systems. In K. Himma and H. Tavani, eds., *The Handbook of Information and Computer Ethics*. Wiley and Sons, New York.

38. Nicki Georghiou and Angus Evans. 2016. *Brexit: the impact on equalities and human rights*. Edinburgh.

39. Hamed Haddadi, Heidi Howard, Amir Chaudhry, Jon Crowcroft, Anil Madhavapeddy, and Richard Mortier. 2015. *Personal Data: Thinking Inside the Box*. London/Cambridge.

40. Marit Hansen, John Morris, Alissa Cooper, et al. Privacy Considerations for Internet Protocols. .

41. Edina Harbinja. 2013. Does the EU Data Protection Regime Protect Post-Mortem Privacy and What Could Be The Potential Alternatives? *SCRIPTed* 10, 1: 19–38.

42. G Iachello and J Hong. 2007. End User Privacy in Human Computer Interaction. *Foundations and Trends in Human Computer Interaction* 1, 1: 1–137.

43. International Telecommunication Union. 2012. *Overview of the Internet of Things*. Geneva.

44. IoT-UK. 2017. *Establishing the Norm: Introduction to IoT Standards*. London.

45. Tom Kirkham, Sandra Winfield, Serge Ravet, and Sampo Kellomaki. 2013. The personal data store approach to personal data security. *IEEE Security and Privacy* 11, 5: 12–19.

46. Max Van Kleek, Daniel Smith, Nigel Shadbolt, and M.c. Schraefel. 2012. A decentralized architecture for consolidating personal information ecosystems: The WebBox. *Proceedings of PIM 2012*.

47. B Koops and R Leenes. 2014. Privacy Regulation Cannot Be Hardcoded. A Critical Comment on the Privacy by Design provision in Data-Protection Law. *International Review of Computers* 28, 2: 159–171.

48. E. Kosta. 2013. Peeking into the cookie jar: the European approach towards the regulation of cookies. *International Journal of Law and Information Technology* 21, 4: 380–406.

49. Kenneth C. Laudon and Kenneth C. 1996. Markets and privacy. *Communications of the ACM* 39, 9: 92–104.

50. Sanna Leppënen and Marika Jokinen. 2003. Daily Routines and Means of Communication in a Smart Home. In R. Harper, ed., *Inside the Smart Home*. Springer Verlag, London, 207–225.

51. L Lessig. 2006. *Code Version 2.0*. Basic Books, New York.

52.    Caroline Leygue, Eamonn Ferguson, Anya Skatova, and Alexa Spence. 2014. Energy Sharing and Energy Feedback: Affective and Behavioral Reactions to Communal Energy Displays. *Frontiers in Energy Research* 2, July: 1–12.

53.    Yong Liu, Jorge Goncalves, Denzil Ferreira, Simo Hosio, and Vassilis Kostakos. 2014. Identity Crisis of Ubicomp?Mapping 15 Years of the Field's Development andParadigm Change. *UbiComp '14*: 75–86.

54.    Ewa Luger, Lachlan Urquhart, Tom Rodden, and Michael Golembewski. 2015. Playing the Legal Card: Using Ideation Cards to Raise Data Protection Issues within the Design Process. *Proceedings of the ACM CHI'15 Conference on Human Factors in Computing Systems*, 457–466.

55.    Derek McAuley. 2014. What is IoT? That is not the Question'. Retrieved from http://iotuk.org.uk/what-is-iot-that-is-not-the-question/ .

56.    Karen McCullagh. 2017. Brexit: potential trade and data implications for digital and 'fintech' industries. *International Data Privacy Law* 7, 1: 3–21.

57.    Richard Mortier, Hamed Haddadi, Tristan Henderson, Derek McAuley, and Jon Crowcroft. 2014. Human-Data Interaction: The Human Face of the Data-Driven Society. *SSRN Electronic Journal*.

58.    Mydex. 2017. Mydex charter. Retrieved from https://pds.mydex.org/mydex-charter.

59.    Patricia A. Norberg, Daniel R. Horne, and David A. Horne. 2007. The Privacy Paradox : Personal Information Disclosure Intentions vers us Behaviors. *The Journal of Consumer Affairs* 41, 1: 100–126.

60.    Out-law. 'General and indiscriminate' data retention laws prohibited, rules EU court. *Out-law.com*. Retrieved 3 March 2017 from https://www.out-law.com/en/articles/2016/december/general-and-indiscriminate-data-retention-laws-prohibited-rules-eu-court/.

61.    Out-Law. 2016. New GDPR guidance: APIs can help businesses meet data portability obligations, says watchdog. *Out-Law.com*. Retrieved 3 March 2017 from https://www.out-law.com/en/articles/2016/december/new-gdpr-guidance-apis-can-help-businesses-meet-data-portability-obligations-says-watchdog/.

62.    OwnCloud. 2017. ownCloud Website. Retrieved from https://owncloud.org.

63.    Eliza Papadopoulou, Alex Stobart, Nick K. Taylor, and M. Howard Williams. 2015. Enabling data subjects to remain data owners. *Smart Innovation, Systems and Technologies*, 239–248.

64.    Article 29 Working Party. 2017. *Guidelines on the Right to Data Portability*. Brussels.

65.    Corien Prins. 2006. Property and Privacy : European Perspectives and the Commodification of our Identity. 5, 223–257. Retrieved 3 March 2017 from https://www.recht.nl/doc/10.Prins.pdf.

66.    Locker Project. 2012. Locker Project. .

67.    J Reidenberg. 1998. Lex Informatica: The Formulation of Policy Rules through Technology. *Texas Law Review* 76: 553.

68.    Joint Committee on Human Rights. 2017. *The Human Rights Implications of Brexit Fifth Report of Session 2016–17*. London.

69.    Tom Rodden and Steve Benford. 2003. The evolution of buildings and implications for the design of ubiquitous domestic environments. *Proceedings of the SIGCHI conference on Human factors in computing systems* 5, 1: 9–16.

70.    Yvonne Rogers. 2006. Moving on from Weiser's Vision of Calm Computing: engaging UbiComp experiences. *Ubicomp '06*: 404–421.

71.    John Rose, Olaf Rehse, and Björn Röber. 2012. The Value of our Digital Identity. *Liberty Global Policy Series*.

72.    K Rose. 2015. Internet of Things: An Overview. *Geneva: Internet Society*.

73. M Satyanarayanan. 2001. Pervasive Computing: Visions and Challenges' Communications. *IEEE Personal Communications* 8, 4: 10–17.

74. Martin Sloan. 2014. Government publishes its review into the midata scheme. *Brodies.com*. Retrieved 3 March 2017 from http://www.brodies.com/blog/government-published-review-midata-scheme/.

75. Daniel J. Solove. 2008. *Understanding Privacy*. Harvard University Press, Cambridge, MA.

76. Sarah Spiekermann and LF Cranor. 2009. Engineering Privacy. *IEEE Transactions on Engineering* 35, 1: 67–82.

77. Peter P. Swire and Yianni Lagos. 2013. Why the Right to Data Portability Likely Reduces Consumer Welfare: Antitrust and Privacy Critique. *SSRN Electronic Journal* 72: 335.

78. Tarryn Ryan and Veronica Scott. 2014. Australia Legislates for Privacy by Design. *The Privacy Advisor (IAPP Website)*.

79. Peter Tolmie, Andy Crabtree, Tom Rodden, and James Colley. 2016. ' This has to be the cats ' - Personal Data Legibility in Networked Sensing Systems. *CSCW 16*, 491–502.

80. Peter Tolmie, James Pycock, Tim Diggins, Allan MacLean, and Alain Karsenty. 2002. Unremarkable computing. *Computer-Human Interaction (CHI) Conference 2002* 1, 1: 399–406.

81. UK Information Commissioner Office. 2017. *Consultation: GDPR Consent Guidance*. Wilmslow.

82. UK Information Commissioner Office. 2017. The right to data portability. Retrieved 3 March 2017 from https://ico.org.uk/for-organisations/data-protection-reform/overview-of-the-gdpr/individuals-rights/the-right-to-data-portability/.

83. Blase Ur, Jaeyeon Jung, and Stuart Schechter. 2014. Intruders Versus Intrusiveness : Teens ' and Parents ' Perspectives on Home-Entryway Surveillance. *UbiComp*, 129–139.

84. Lachlan Urquhart. 2017. Ethical Dimensions of User Centric Regulation. *Proceedings of CEPE/ETHICOMP 2017*.

85. Lachlan Urquhart. 2017. Towards User Centric Regulation. .

86. Lachlan Urquhart and Tom Rodden. 2017. New directions in information technology law: learning from human–computer interaction. *International Review of Law, Computers & Technology* 31, 2: 1–19.

87. L Urquhart and T Rodden. 2016. A Legal Turn in HCI: Towards Regulation by Design for the Internet of Things. *Social Science Research Network*.

88. Aysem Diker Vanberg and Mehmet Bilal Ünver. 2017. The right to data portability in the GDPR and EU competition law: odd couple or dynamic duo? *European Journal of Law and Technology* 8, 1: 1–22.

89. Sih Yuliana Wahyuningtyas. 2015. Interoperability for data portability between social networking sites (SNS): the interplay between EC software copyright and competition law. *Queen Mary Journal of Intellectual Property* 5, 1: 46–67.

90. Coral Walker and Coral Walker. 2015. Personal Data Lake With Data Gravity Pull Personal Data Lake With Data Gravity Pull. August.

91. Mark Walport. 2014. *The Internet of Things: making the most of the Second Digital Revolution*. London.

92. Mark Weiser. 1993. Some computer science issues in ubiquitous computing. *Communications of the ACM* 36, 7: 75–84.

93. Mark Weiser and John Seely Brown. 1996. *The coming age of calm technology*. Palo Alto.

94. Alan F. Westin. 1968. Privacy and Freedom. *American Sociological Review* 33, 1: 173.
95. C Wilson. 2015. Smart Homes and Their Users: Analysis and Key Challenges. *Personal and Ubiquitous Computing* 19: 463–476.
96. Rebecca Wong and Joseph Savirimuthu. 2008. All or Nothing: This is the Question? The Application of Art. 3(2) Data Protection Directive 95/46/EC to the Internet. *J. Marshall J. Computer and Info Law* 25: 24.
97. World Economic Forum and Boston Consulting Group. 2012. Rethinking Personal Data : Strengthening Trust. May: 1–35.
98. G. Zanfir. 2012. The right to Data portability in the context of the EU data protection reform. *International Data Privacy Law* 2, 3: 149–162.